\documentclass[pra,onecolumn,showpacs,nofootinbib,preprint,floatfix]{revtex4}

\usepackage{amsmath,amsfonts,amssymb,bm}
\usepackage{dcolumn}
\usepackage[final]{graphicx}
\usepackage{bm}
\usepackage{comment}
\usepackage{dsfont}
\usepackage{float}
\usepackage{rotating}
\usepackage{leftidx}
\usepackage{bbold}

\def\beq{\begin{equation}}
\def\eeq{\end{equation}}
\def\beqn{\begin{eqnarray}}
\def\eeqn{\end{eqnarray}}
\def\souligne#1{$\underline{\smash{ \hbox{#1}}}$}

\def\brho {\mbox{\boldmath $\rho$}}
\def\r {{\bf r}}

\def\A {{\bf A}}
\def\B {{\bf B}}
\def\Q {{\bf Q}}

\def\u {{\bf u}}
\def\v {{\bf v}}

\def\1 {{\bf 1}}
\def\0 {{\bf 0}}

\def\f  {{\bf f}}
\def\g {{\bf g}}

\def\C {{\bf C}}

\def\bcalL {\mbox{\boldmath $\mathcal L$}}

\begin{document}

\title{Many-body excitation spectra of trapped bosons
with general interaction by linear response}

\author{Ofir E. Alon}

\affiliation{Department of Physics, University of Haifa at Oranim, Tivon 36006, Israel}

\date{\today}

\begin{abstract}
The linear-response theory
of the multiconfigurational time-dependent Hartree for bosons method
for computing many-body excitations of trapped Bose-Einstein condensates
[Phys. Rev. A {\bf 88}, 023606 (2013)]
is implemented for systems with general interparticle interaction.
Illustrative numerical examples for repulsive and attractive bosons are provided.
The many-body linear-response theory identifies the excitations not unraveled
within Bogoliubov--de Gennes equations.
The theory is herewith benchmarked against the exactly-solvable 
one-dimensional harmonic-interaction model.
As a complementary result, we represent the 
theory in a compact block-diagonal form,
opening up thereby an avenue for treating larger systems. 
\end{abstract}


\maketitle 

\section{Introduction}\label{sec}

The standard and most popular avenue to compute the dynamics (and excitations) 
of a Bose-Einstein condensate (BEC) is the Gross-Pitaevskii equation,
which assumes all bosons to occupy a single one-particle state \cite{Book_Pitaevskii,Book_Leggett,Book_Pethick}.
Clearly, whenever the system under 
investigation is not fully condensed 
one has to go beyond Gross-Pitaevskii theory in order to faithfully account for the system's dynamics,
a matter which is well documented in the literature,
see, e.g., the recent book \cite{Nick_book_2013}
and references therein.

The usual way to account for excitations in a BEC is to build them atop the Gross-Pitaevskii ground state,
taking a particle out from the condensate to an excited one-particle state.
Formally, this leads to the Bogoliubov--de Gennes equations
which often are also referred to as linear response
of the Gross-Pitaevskii equation \cite{BdG,Book_Pitaevskii,Book_Leggett,Book_Pethick,Nick_book_2013,
LR_GP_Ruprecht,Esry,C_Gardiner,Castin_Dum}. 
It turns out,
as we have recently shown \cite{LR_MCTDHB}, 
that even if the ground state is well described
by the fully-condensed Gross-Pitaevskii wavefunction,
many excitations are missing by this standard treatment.
For fully-fragmented ground states \cite{BMF},
more excitations appear \cite{LR_BMF}.
Thus, to describe systems whose ground states are 
not fully condensed and, importantly, 
to identify new classes of excitations which cannot 
be resolved by the standard tools,
suitable methods are in need.

Consider $N$ trapped bosons,
interacting by a generic interparticle interaction.
The many-particle Hamiltonian is written as follows
\beq\label{Ham}
 \hat H(\r_1,\ldots,\r_N) = \sum_{j=1}^N \hat h(\r_j) + \sum_{k>j=1}^N \hat W(\r_j,\r_k).
\eeq
Here $\hat h(\r)$
is the one-body Hamiltonian which consists of
kinetic and potential (trap) terms
and $\hat W(\r,\r')$ is a generic interparticle interaction which
is symmetric to permutation of the particles' coordinates.
In the familiar case for ultracold bosonic atoms
it takes the form $\hat W(\r,\r') = \lambda_0 \delta(\r-\r')$
where the interaction parameter $\lambda_0$
is proportional to the s-wave scattering length.

We recall that by approximating the system's time-dependent wavefunction by the product state
\beq\label{GP_Psi}
\Psi_{\mathrm{GP}}(\r_1,\ldots,\r_n;t) = \phi(\r_1,t) \cdots \phi(\r_N,t) \equiv |N;t\rangle
\eeq
and utilizing the Dirac-Frenkel variational principle, 
the Gross-Pitaevskii equation,
$i \dot\phi(\r,t) = [\hat h(\r) + \lambda |\phi(\r,t)|^2] \phi(\r,t)$,
where $\lambda = \lambda_0 (N-1)$,
is obtained.
 
The linear-response theory atop the Gross-Pitaevskii mean-field wavefunction (\ref{GP_Psi})
is formally obtained by linearizing the Gross-Pitaevskii equation
around the ground-state solution.
It results in the Bogoliubov--de Gennes equations which take on the matrix form
\beq\label{BDG_eq}
\bcalL_{\mathrm{BdG}} 
\left(\!\!\begin{array}{c} 
u^k \\ 
v^k \\
\end{array}\!\!\right) =
\omega_k 
\left(\!\!\begin{array}{c} 
u^k \\ 
v^k \\
\end{array}\!\!\right), \qquad
\bcalL_{\mathrm{BdG}} = 
\left(\!\begin{array}{cc} \hat h 
+2\lambda|\phi_0|^2-\mu & \lambda(\phi_0)^2\\
-\lambda(\phi_0^\ast)^{2} &  -(\hat h^\ast +2\lambda|\phi_0|^2-\mu)
 \end{array}\!\right).
\eeq
The linear-response matrix $\bcalL_{\mathrm{BdG}}$ 
depends explicitly on the ground-state
orbital $\phi_0$.
$\mu$ is the chemical potential.
The excitation spectrum $\omega_k$ of the BEC
as well as the so-called response amplitudes $u^k$ and $v^k$
are obtained by solving the eigenvalue system (\ref{BDG_eq}).

\section{Many-body theory}\label{Theory}

In many situations as mentioned above 
the ground state of the BEC cannot be described well 
by the wavefunction Eq.~(\ref{GP_Psi}).
Importantly, even when
the ground state is well approximated by the Gross-Pitaevskii wavefunction (\ref{GP_Psi}),
the standard linear-response atop misses many excitations \cite{LR_MCTDHB},
also see the benchmarks in Sec.~\ref{num_ben} below.
The natural idea was to use 
a more extended ansatz for the
system's wavefunction, 
and then to perform 
linear response atop \cite{LR_MCTDHB,ULR_MCTDH}.
We recall that linear response atop
the exact ground state gives rise to
the exact many-body excitation spectrum,
see, e.g., \cite{Fetter_book,JCP_exact_LR}.

Let us briefly describe the underlying many-body theory for BECs used in the present work
before we proceed to its linear response.
In the multiconfigurational time-dependent Hartree for bosons (MCTDHB) 
method \cite{MCTDHB1,MCTDHB2} the bosons
are allowed to occupy not one but $j=1,\ldots,M$ one-particle functions (modes) $\phi_j$.
The many-body wavefunction is assembled 
by distributing the $N$ bosons over the $M$ one-particle functions
\beq\label{MCTDHB_Psi}
 \Psi(t) = \sum_{\vec n} C_{\vec n}(t) |\vec n;t\rangle,
\eeq
where $C_{\vec n}(t)$ are the expansion coefficients
and $|\vec n;t\rangle$ are permanents (Fock states)
with $\vec n \equiv (n_1,\ldots,n_M)$, $n_1 + \ldots + n_M = N$. 
Utilizing the Dirac-Frenkel variational principle,
the one-particle functions $\phi_j$ as well as the expansion 
coefficients $C_{\vec n}$ are determined self-consistently.
This leads to a system of coupled equations 
which has been coined
in the literature the MCTDHB method \cite{MCTDHB1,MCTDHB2}.
The MCTDHB method has been used
for unveiling many-body phenomena with repulsive and attractive BECs 
in one-dimensional setups, see, e.g., \cite{BJJ,swift,Julian_OCT,Wave_Chaos_depletion},
and benchmarked against an exactly-solvable model \cite{HIM_Axel}. 
Most recently, MCTDHB was extended to two and three spatial dimensions,
and employed to establish the mechanism of fragmentation and generic regimes of dynamics 
in repulsive BECs with strong, finite-range interparticle interactions \cite{Customized,dynamics_long}.

The successes of the MCTDHB method to accurately compute 
many-body out-of-equilibrium dynamics of BECs stem
from the employment of the self-consistent (time-adaptive) multiconfigurational wavefunction (\ref{MCTDHB_Psi}).
The wavefunction (\ref{MCTDHB_Psi}) contains
a substantially larger number of variational parameters
(i.e., modes, and Fock states and their expansion coefficients)
in comparison with the standard Gross-Pitaevskii wavefunction (\ref{GP_Psi}).
Clearly, if the obtained many-body dynamics is accurate
then MCTDHB manages to resolve the
excitation spectra of BECs.
This has motivated us to pursue its linear response,
as a venue to research on the many-body level excitations
of trapped BECs directly, i.e., without propagation.

The derivation of the linear-response (LR) theory atop the wavefunction (\ref{MCTDHB_Psi}) 
is rather lengthly but otherwise straightforward \cite{LR_MCTDHB,ULR_MCTDH}.
We will not repeat it here and 
begin from the final result for the
resulting LR-MCTDHB theory,
which takes on the form of the eigenvalue equation \cite{LR_MCTDHB,ULR_MCTDH}
\beq\label{LR_MCTDHB}
\bcalL 
\left(\!\!\begin{array}{c} 
\u^k \\ 
\v^k \\
\C_u^k \\ 
\C_v^k \\
\end{array}\!\!\right) =
\omega_k 
\left(\!\!\begin{array}{c} 
\u^k \\ 
\v^k \\
\C_u^k \\ 
\C_v^k \\
\end{array}\!\!\right).
\eeq
The linear-response matrix $\bcalL$ of the many-boson 
wavefunction $\Psi$ is more involved than the
Bogoliubov--de Gennes linear-response matrix (\ref{BDG_eq}).
We will discuss its structure shortly.
Physically, the response amplitudes of all modes,
$\u^k$ and $\v^k$,
and of all expansion coefficients,
$\C_u^k$ and $\C_v^k$,
combine to give the many-body excitation spectrum $\omega_k$.
For comparison,
in the Bogoliubov--de Gennes linear-response matrix (\ref{BDG_eq})
there is only a single block representing the
sole one-particle function 
used to describe the BEC within Gross-Pitaevskii theory.
In Ref.~\cite{LR_MCTDHB}
we have successfully managed to explicitly construct $\bcalL$ for bosons
interacting by contact potential and obtained the
many-body excitation spectrum.

There is another way to group the linear-response matrix which we are now going to exploit.
Namely, first to list the orbitals' and coefficients' `u' blocks and then the respective `v' blocks.
In this way each of the new blocks has the same dimension, see below.
The spectrum, of course, does not change.
Hence, reshuffling the blocks of $\bcalL$ the final result can be written as
\beq\label{LR_matrix_diag_2}
\bcalL
\begin{pmatrix} 
 \u^k \\
 \C_u^k \\ \hline
 \v^k \\
 \C_v^k \\
\end{pmatrix} =
 w_k
\begin{pmatrix} 
 \u^k \\
 \C_u^k \\ \hline
 \v^k \\
 \C_v^k \\
\end{pmatrix}, \qquad
\bcalL =
\left(\begin{array}{c|c} 
\bcalL^u & \bcalL^v \\ \hline
-{(\bcalL^v)}^\ast & -{(\bcalL^u)}^\ast \\
 \end{array}\right),
\eeq
where details of the blocks of $\bcalL$ are collected in \cite{L_note}.

The above general relation between the `u' and `v' blocks of $\bcalL$
is appealing
since we may mix them and eventually block diagonalize $\bcalL$.
For this,
consider the transformation
\beq\label{app_TRANS}
 \Q = \frac{1}{\sqrt{2}}\left(\begin{array}{cr} 
\1 & \1 \theta \\
\1 & -\1 \theta \\
\end{array}\right),
\eeq
where $\theta$ is the operation of complex conjugation, for example, $\theta \v^k = (\v^k)^\ast$.
It is not difficult to show that $\Q$ block diagonalizes $\bcalL$;
Consult the appendix for additional details.
The final result reads
\beqn\label{Block_final}
& &
\bcalL_f^{(2)}
\begin{pmatrix} 
 \f^k \\
 \C_f^k \\
\end{pmatrix} =
 w^2_k
\begin{pmatrix} 
 \f^k \\
 \C_f^k \\
\end{pmatrix}, \qquad
\bcalL_f^{(2)} = (\bcalL^u-\bcalL^v\theta) (\bcalL^u+\bcalL^u\theta), \nonumber \\
& & 
\bcalL_g^{(2)}
\begin{pmatrix} 
 \g^k \\
 \C_g^k \\
\end{pmatrix} =
 w^2_k
\begin{pmatrix} 
 \g^k \\
 \C_g^k \\
\end{pmatrix}, \qquad
\bcalL_g^{(2)} = (\bcalL^u+\bcalL^u\theta) (\bcalL^u-\bcalL^v\theta),
\eeqn 
with the relations between the eigenvectors' blocks
$\f^k,\g^k = \frac{1}{\sqrt{2}} [\u^k \pm ({\v^k})^\ast]$ and
$\C^k_{f},\C^k_{g} = \frac{1}{\sqrt{2}} [\C_u^k \pm (\C_v^k)^\ast]$.

Eq.~(\ref{Block_final}) is the main result of this work on the theory side;
We compactly represent the LR-MCTDHB theory in a block-diagonal form.
Because the resulting (square of the) excitation spectrum 
is bound from below,
this will open up an avenue for treating larger systems.
First, this in particular would allow one to use standard diagonalization
techniques for matrices with bound spectra.
Here, Eq.~(\ref{Block_final}) defines the basic operation
of matrix-to-vector multiplication used
in such techniques.
Second, it reduces the size of the linear-response matrix to half its size.
Furthermore, since $\Q$ in Eq.~(\ref{app_TRANS}) is a complex transformation,
the many-body wavefunction atop which 
the many-body linear response is
performed can be a complex quantity.
Such a construction generalizes the
block-diagonalization treatment of the Bogoliubov--de Gennes equations,
see, e.g., \cite{BdG_B1_LS,BdG_B2_Ronen}.
This concludes our block diagonalization of the LR-MCTDHB theory.

\section{Illustrative numerical examples and benchmarks}\label{num_ben}

In the following section we would like to report the implementation
and application of LR-MCTDHB with general interparticle interaction.
As an illustrative system we have chosen the harmonic-interaction model \cite{HIM1,HIM2}.
This is an analytically-solvable model, yet, it is not at all trivial to be treated numerically.
First, because the interaction is non-contact the
construction of various matrix elements for computing the ground-state wavefunction and
its response matrix are much more involved.
We have now successfully coped with this task.
Second, the solution of the problem on the computer is done in the laboratory frame,
where all the bosons are indistinguishable.
This is unlike the analytical solution which
exploits the separability of the 
center-of-mass and relative coordinates' degrees-of-freedom.
As mentioned above,
the harmonic-interaction model and
a time-dependent extension of which have been used to benchmark MCTDHB \cite{HIM_Axel}.
Hence, we expect the model to be instrumental in benchmarking the
excitation spectrum computed by the many-body linear-response theory.

In Eq.~(\ref{Ham}) the one-body Hamiltonian 
is now $\hat h(x) = -\frac{1}{2} \frac{\partial^2}{\partial x^2} + \frac{1}{2} \Omega^2 x^2$.
The two-body interaction reads $W(x-x')=K(x-x')^2$,
representing thereby long-range interaction.
The parameter $K<0$ ($K>0$) indicates repulsion (attraction)
between the bosons.
In what follows we set without loss of generality $\Omega=1$.
The exact excitation energies of the one-dimensional harmonic-interaction model
are known and given by \cite{HIM1,HIM2}
\beq\label{HIM_spec}
 \omega[n_{CM}, n_{rel}] = n_{CM} + n_{rel} \delta_N, \qquad \delta_N=\sqrt{1 + 2NK},
\eeq
with the center-of-mass ($CM$) and relative coordinates' ($rel$) quantum numbers 
$n_{CM}=1,2,3,\ldots$ and $n_{rel}=2,3,\dots$.

We consider $N=1000$ weakly-interacting repulsive bosons ($K=-0.0001$)
and compare in Table~\ref{T1} below the LR-MCTDHB results
with the analytical formula (\ref{HIM_spec}).
The many-body theory improves as one would expect the accuracy of excitations unveiled
by Bogoliubov--de Gennes theory.
More important,
it unravels additional excitations.
We see the capability of the LR-MCTDHB theory to describe numerically-exactly 
the center-of-mass and relative coordinates' excitations.
We stress again that the computation is done in the laboratory frame,
where all bosons are equivalent, i.e.,
the separability to center-of-mass and relative coordinates,
which is special for harmonic traps, 
is not exploited.

\begin{table}
\begin{tabular}{|c|c|c|c|c|}
\hline                 &    M=1              &    M=2            &      Exact analytical       & $n_{CM}$, $n_{rel}$ \\ \hline
$E_{GS}$           &   447.26\souligne{949371}   &   447.2663819\souligne{4}   &  447.26638190                     & 0, 0  \\ \hline
$\omega_1$        &  1.00000000       &   1.00000000      &  1.00000000        & 1, 0  \\ \hline
$\omega_2$        &  1.78\souligne{907797}       &   1.788854\souligne{43}      &  1.78885438        & 0, 2  \\ \hline
$\omega_3$        &        n/a            &   2.0000\souligne{4476}      &  2.00000000        & 2, 0  \\ \hline
$\omega_4$        &  2.683\souligne{61696}       &   2.683281\souligne{68}      &  2.68328157        & 0, 3  \\ \hline
$\omega_5$        &        n/a            &   2.7888\souligne{8891}      &  2.78885438        & 1, 2  \\ \hline
$\omega_6$        &        n/a            &   3.0000\souligne{7751}      &  3.00000000        & 3, 0  \\ \hline
$\omega_7$        &  3.57\souligne{815595}       &   3.5777\souligne{1028}      &  3.57770876        & 0, 4  \\ \hline
$\omega_8$        &        n/a            &   3.683\souligne{97387}      &  3.68328157        & 1, 3  \\ \hline
\end{tabular}
\caption{Spectrum of the one-dimensional harmonic-interaction model
with $N=1000$ bosons and repulsion $K=-0.0001$.
Comparisons of LR-MCTDHB and the exact results for the ground, $E_{GS}$, and excited states, 
$\omega_k=E_k-E_{GS}$. 
The last column assigns
the excitations in terms
of center-of-mass and relative coordinates' quantum numbers.
Some excitations are first uncovered at the $M=2$
level of theory, i.e., they are not available (n/a) within Bogoliubov--de Gennes theory ($M=1$).
Convergence with $M$ to the exact results is clearly seen.
All quantities are dimensionless.}
\label{T1}
\end{table}

We now move to attractive interaction and compute the excitation spectrum
of $N=1000$ bosons with $K=+0.0001$.
The results are collected in Table~\ref{T2}
and show, as above, the capability of LR-MCTDHB
to uncover the missing excitations and to numerically-converge to the analytical results.
When comparing the repulsive and attractive systems,
please note the interchange of order of some center-of-mass and relative coordinates' excitations.

\begin{table}
\begin{tabular}{|c|c|c|c|c|}
\hline                 &    M=1                &    M=2                &    Exact analytical      &   $n_{CM}$, $n_{rel}$ \\ \hline
$E_{GS}$           &    547.67\souligne{691206}     &  547.6748349\souligne{7}      &    547.67483495         &               0, 0  \\ \hline
$\omega_1$        &       1.00000000     &     1.00000000     &        1.00000000         &              1, 0  \\ \hline
$\omega_2$        &        n/a               &     2.0000\souligne{3648}     &        2.00000000        &               2, 0  \\ \hline
$\omega_3$        &        2.190\souligne{70765}    &     2.1908902\souligne{6}     &        2.19089023         &               0, 2  \\ \hline
$\omega_4$        &        n/a               &     3.0000\souligne{6435}     &        3.00000000        &               3, 0  \\ \hline
$\omega_5$        &        n/a               &     3.190\souligne{91687}     &        3.19089023        &               1, 2  \\ \hline
$\omega_6$        &        3.286\souligne{06147}     &     3.286335\souligne{42}     &        3.28633535        &               0, 3  \\ \hline
$\omega_7$        &        n/a                &     4.00\souligne{129579}    &         4.00000000        &               4, 0  \\ \hline
$\omega_8$        &        n/a                &     4.28\souligne{580642}    &         4.28633535       &                1, 3  \\ \hline
\end{tabular}
\caption{Same as Table~\ref{T1} but for $N=1000$ bosons 
with attraction $K=+0.0001$.
Some excitations are first uncovered at the $M=2$
level of theory, i.e., they are not available (n/a) within Bogoliubov--de Gennes theory ($M=1$).
Convergence with $M$ to the exact results is clearly seen.
Note the interchange of order of some center-of-mass and relative coordinates' 
excitations in comparison with
the repulsive system.
All quantities are dimensionless.}
\label{T2}
\end{table}

\section{Concluding remarks}\label{con}

The linear-response theory
of the multiconfigurational time-dependent Hartree for bosons method
for computing many-body excitations of trapped Bose-Einstein condensates
has been implemented for systems with general interparticle interaction.
This allows us to investigate the excitation spectrum
of interacting bosons with, for instance, long-range interaction.
As illustrative examples we considered, separately, repulsive and
attractive bosons within the one-dimensional harmonic-interaction model.
The many-body linear-response theory is capable of identifying all excitations,
including the excitations which are not unraveled
within Bogoliubov--de Gennes equations.
The results of the present work serve to benchmark the LR-MCTDHB method.

As a complementary result, we compactly represent the 
theory in a block-diagonal form.
This will open up an avenue for treating larger systems.
We expect the LR-MCTDHB theory and its implementation
for general interparticle interaction to provide
an important probe into the many-body excitation involved
in the out-of-equilibrium dynamics of trapped BECs.

\section*{Acknowledgements}

I am grateful to Alexej Streltsov and Lorenz Cederbaum for stimulating and continuous discussions. 

\appendix*\label{Appendix}

\section{Block diagonalization by a complex transformation}\label{com_appen}

Consider the eigenvalue system
\beq\label{app_general_EV}
\left(\begin{array}{cc} 
\A & \B \\
-\B^\ast & -\A^\ast \\
\end{array}\right)
\begin{pmatrix} 
 \u \\
 \v \\
\end{pmatrix} =
\omega
\begin{pmatrix} 
 \u \\
 \v \\
\end{pmatrix},
\eeq
where 
$\A,\B$ and $\u,\v$ are the blocks
of a square matrix and its eigenvector, respectively.
The eigenvalue $\omega$ is assumed to be real.

Let us examine the transformation matrix
$\Q = \frac{1}{\sqrt{2}}\left(\begin{array}{cr} 
\1 & \1 \theta \\
\1 & -\1 \theta \\
\end{array}\right)$,
Eq.~(\ref{app_TRANS}) of the main text.
Interestingly, the transformation matrix $\Q$ is neither a unitary nor an anti-unitary operator.
Its inverse exists and reads
\beq\label{app_TRANS_m1}
 \Q^{-1} = \frac{1}{\sqrt{2}}\left(\begin{array}{lc} 
\1 & \1 \\
\1 \theta & -\1 \theta \\
\end{array}\right), \qquad  \qquad \Q \Q^{-1} = \Q^{-1} \Q = 
\left(\begin{array}{cc} 
\1 & \0 \\
\0 & \1 \\
\end{array}\right).
\eeq
Multiplying with $\Q$ from the left on both sides of Eq.~(\ref{app_general_EV}) we find
\beq\label{app_Q_transform}
\left\{\Q  \left(\begin{array}{cc} 
\A & \B \\
-\B^\ast & -\A^\ast \\
\end{array}\right)  \Q^{-1} \right\}
\left\{\Q  \begin{pmatrix} 
 \u \\
 \v \\
\end{pmatrix} \right\} =
\left(\begin{array}{cc} 
\0 & \A-\B\theta \\
\A+\B\theta &  \0 \\
\end{array}\right) 
\begin{pmatrix} 
 \f \\
 \g \\
\end{pmatrix} =
\omega
\begin{pmatrix} 
 \f \\
 \g \\
\end{pmatrix},
\eeq
where 
$\begin{pmatrix} 
 \f \\
 \g \\
\end{pmatrix} =
\begin{pmatrix} 
  \frac{1}{\sqrt{2}}(\u + \v^\ast) \\
  \frac{1}{\sqrt{2}}(\u - \v^\ast) \\
\end{pmatrix}$.
Multiplying now Eq.~(\ref{app_Q_transform}) from the left with the transformed matrix we obtain the 
desired result in a block-diagonal form
\beq
\left(\begin{array}{cc} 
(\A-\B\theta) (\A+\B\theta) & \0 \\
\0 & (\A+\B\theta) (\A-\B\theta)\\
\end{array}\right) 
\begin{pmatrix} 
 \f \\
 \g \\
\end{pmatrix} =
\omega^2
\begin{pmatrix} 
 \f \\
 \g \\
\end{pmatrix}.
\eeq
This concludes our derivation.

Using the operation of complex conjugation $\theta$ above becomes redundant if
$\A$ and $\B$ are real quantities.
In this case, which is usually considered within Bogoliubov--de Gennes equations,
see, e.g., \cite{BdG_B1_LS,BdG_B2_Ronen},
$\Q$ becomes a unitary operator.


\begin{thebibliography}{99}

\bibitem{Book_Pitaevskii} L. Pitaevskii and S. Stringari, 
                                   {\it Bose-Einstein Condensation} (Oxford University Press, Oxford, 2003).

\bibitem{Book_Leggett} A. J. Leggett, 
                       {\it Quantum Liquids: Bose condensation and Cooper pairing in condensed matter systems}
                       (Oxford University Press, Oxford, 2006).

\bibitem{Book_Pethick} C. J. Pethick and H. Smith, 
                                {\it Bose-Einstein Condensation in Dilute Gases}, 2nd ed. 
                                (Cambridge University Press, Cambridge, England, 2008).

\bibitem{Nick_book_2013} {\it Quantum Gases: Finite Temperature and 
Non-Equilibrium Dynamics (Vol. 1 Cold Atoms Series)}, 
edited by N. P. Proukakis, S. A. Gardiner, M. J. Davis, 
and M. H. Szymanska (Imperial College Press, London, 2013).

\bibitem{BdG} N. N. Bogoliubov,
                      J. Phys. USSR {\bf 11}, 23 (1947)
                      (reprinted in D. Pines, {\it The Many-body Problem}, 
                      W. A. Benjamin, New York, 1961).

\bibitem{LR_GP_Ruprecht} P. A. Ruprecht, M. Edwards, K. Burnett, and C. W. Clark, 
                         Phys. Rev. A {\bf 54}, 4178 (1996).

\bibitem{Esry} B. D. Esry,
                     Phys. Rev. A {\bf 55}, 1147 (1997).

\bibitem{C_Gardiner} C. W. Gardiner, 
 Phys. Rev. A {\bf 56}, 1414 (1997).

\bibitem{Castin_Dum} Y. Castin and R. Dum,
                     Phys. Rev. A {\bf 57}, 3008 (1998). 

\bibitem{LR_MCTDHB} J. Grond, A. I. Streltsov, A. U. J. Lode, K. Sakmann, L. S. Cederbaum, and O. E. Alon, 
Phys. Rev. A {\bf 88}, 023606 (2013).

\bibitem{BMF} L. S. Cederbaum and A. I. Streltsov, 
Phys. Lett. A {\bf 318}, 564 (2003).

\bibitem{LR_BMF} J. Grond, A. I. Streltsov, L. S. Cederbaum, and O. E. Alon, 
 Phys. Rev. A {\bf 86}, 063607 (2012).

\bibitem{ULR_MCTDH} O. E. Alon, A. I. Streltsov, and L. S. Cederbaum, 
                                J. Chem. Phys. {\bf 140}, 034108 (2014).

\bibitem{Fetter_book} A. L. Fetter and J. D. A Walecka, 
                                {\it Quantum Theory of Many-Particle Systems} 
                                (McGraw-Hill, New York, 1971).

\bibitem{JCP_exact_LR} J. Olsen and P. J{\o}rgensen,
                                  J. Chem. Phys. {\bf 82}, 3235 (1985).

\bibitem{MCTDHB1} A. I. Streltsov, O. E. Alon, and L. S. Cederbaum,
                  Phys. Rev. Lett. {\bf 99}, 030402 (2007).

\bibitem{MCTDHB2} O. E. Alon, A. I. Streltsov, and L. S. Cederbaum,
                  Phys. Rev. A {\bf 77}, 033613 (2008).

\bibitem{BJJ} K. Sakmann, A. I. Streltsov, O. E. Alon, and L. S. Cederbaum,
              Phys. Rev. Lett. {\bf 103}, 220601 (2009).

\bibitem{swift}  A. I. Streltsov, O. E. Alon, and L. S. Cederbaum,
                       Phys. Rev. Lett. {\bf 106}, 240401 (2011).

\bibitem{Julian_OCT} J. Grond, J. Schmiedmayer, and U. Hohenester, 
                               Phys. Rev. A {\bf 79}, 021603(R) (2009). 

\bibitem{Wave_Chaos_depletion} I. B$\check{\mathrm r}$ezinov\'a, A. U. J. Lode, 
               A. I. Streltsov, O. E. Alon, L. S. Cederbaum, and J. Burgd\"orfer,
                                     Phys. Rev. A {\bf 86}, 013630 (2012). 

\bibitem{HIM_Axel} A. U. J. Lode, K. Sakmann, O. E. Alon, L. S. Cederbaum, and A. I. Streltsov, 
            Phys. Rev. A {\bf 86}, 063606 (2012).

\bibitem{Customized} A. I. Streltsov,
      Phys. Rev. A {\bf 88}, 041602(R) (2013). 

\bibitem{dynamics_long} O. I. Streltsova, O. E. Alon, L. S. Cederbaum, and A. I. Streltsov, 
      Phys. Rev. A {\bf 89}, 061602(R) (2014). 

\bibitem{L_note} The blocks of the linear-response matrix (\ref{LR_matrix_diag_2}) are given by
$$
\bcalL^u =
\left(\begin{array}{cc} 
{\mathbf P} \brho^{+\frac{1}{2}} \bcalL_{oo}^u \brho^{+\frac{1}{2}} {\mathbf P}  \ & 
\ {\mathbf P} \brho^{+\frac{1}{2}} \bcalL_{oc}^u {\mathbf P}_C  \\ 
{\mathbf P}_C  (\bcalL_{oc}^u)^\dag \brho^{+\frac{1}{2}} {\mathbf P} \ & 
\ {\mathbf P}_C   (\cdot)^{\hat H - \varepsilon}  {\mathbf P}_C  \\
 \end{array}\right),
\ \ \quad
\bcalL^v =
\left(\begin{array}{cc} 
{\mathbf P} \brho^{+\frac{1}{2}} \bcalL_{oo}^v (\brho^\ast)^{+\frac{1}{2}} {\mathbf P}^\ast  \ & 
\ {\mathbf P} \brho^{+\frac{1}{2}} \bcalL_{oc}^v {\mathbf P}_C^\ast  \\ 
{\mathbf P}_C  (\bcalL_{oc}^v)^t (\brho^\ast)^{+\frac{1}{2}} {\mathbf P}^\ast \ & 
\ \0  \\
 \end{array}\right).
$$
The meaning of the ingredients is as follows:
${\mathbf P}$ is a projector matrix on the subspace of one-particle functions
orthogonal to the ground-state modes $\phi_j$;
${\mathbf P}_C$ is a projector matrix on the subspace of permanents' coefficients
orthogonal to the ground-state coefficients $C_{\vec n}$;
$\brho$ is the reduce one-particle density matrix of the ground state;
and $ (\cdot)^{\hat H - \varepsilon}$ marks the operation of the Hamiltonian in Fock space
utilizing mapping of permanents \cite{Mapping} and $\varepsilon$ is the ground-state energy.
Finally, the sub-blocks $\bcalL_{oo}^u$, $\bcalL_{oo}^v$ and $\bcalL_{oc}^u$, $\bcalL_{oc}^v$
collect, respectively, the couplings between the modes and the couplings between
the modes and coefficients.
Together with further details, 
they are given in \cite{LR_MCTDHB,ULR_MCTDH}.

\bibitem{BdG_B1_LS} C. Huepe, L. S. Tuckerman, S. M\'etens, and M. E. Brachet,
                          Phys. Rev. A {\bf 68}, 023609 (2003).

\bibitem{BdG_B2_Ronen} S. Ronen, D. C. E. Bortolotti, and J. L. Bohn,
                                   Phys. Rev. A {\bf 74}, 013623 (2006).

\bibitem{HIM1} L. Cohen and C. Lee, 
                      J. Math. Phys. {\bf 26}, 3105 (1985). 

\bibitem{HIM2} J. Yan, 
                      J. Stat. Phys. {\bf 113}, 623 (2003).

\bibitem{Mapping} A. I. Streltsov, O. E. Alon, and L. S. Cederbaum, 
                           Phys. Rev. A {\bf 81}, 022124 (2010). 

\end{thebibliography}
\end{document}